%% file: ms_submitted211207.tex
\documentclass[12pt,preprint,usenatbib]{emulateapj}

\newcommand{\muhz}{$\mu$Hz}
\newcommand{\numax}{$\nu_{\mathrm{max}}$}
\newcommand{\wire}{{\it{WIRE}}}

\shorttitle{Oscillating K~giants with the \wire\ satellite}
\shortauthors{Stello et al.}

\received{2007 December 5}
\begin{document}



\title{Oscillating K~giants with the \wire\ satellite: \\
    determination of their asteroseismic masses}



\author{D. Stello\altaffilmark{1}, H. Bruntt\altaffilmark{1}, H. Preston\altaffilmark{2}, and D. Buzasi\altaffilmark{2}}



\altaffiltext{1}{School of Physics, University of Sydney, NSW 2006,
  Australia; stello@physics.usyd.edu.au.}
\altaffiltext{2}{US Air Force Academy, Colorado Springs, CO 80840, USA.}


\begin{abstract}
Mass estimates of K~giants are generally very uncertain. 
Traditionally, stellar masses of single field stars are determined by
comparing their location in the Hertzsprung-Russell diagram with stellar 
evolutionary models.
Applying an additional method to determine the mass is therefore of
significant interest for understanding stellar evolution.
We present the time series analysis of 11 K~giants recently observed 
with the \wire\ satellite. 
With this comprehensive sample, we report the first confirmation that the 
characteristic acoustic frequency, \numax, 
can be predicted for K~giants by scaling from the solar acoustic cut-off
frequency. 
We are further able to utilize our measurements of \numax\ to determine an 
asteroseismic mass for each star with a lower uncertainty compared to
the traditional method, for most stars in our sample. 
This indicates good prospects for the application of our method on the vast
amounts of data that will soon come from the COROT and Kepler space
missions.

\end{abstract}



\section{Introduction}
According to theoretical calculations of stellar evolution, stars of a
large mass range, corresponding to main-sequence spectral classes from late
B to K, all end up in roughly the same part of the Hertzsprung-Russell (H-R)
diagram when they evolve to become red giants with spectral classes from
late G to M. Here, the evolution tracks are closely spaced,
which makes it difficult to estimate the mass, and hence to determine the
progenitor of a red giant star from its position in the H-R diagram.

The investigation of stellar oscillations (asteroseismology), in particular
solar-like oscillations, 
provides a unique tool to probe the interior and hence also 
the mass of stars. 
Much excitement therefore followed the first clear evidence of solar-like
oscillations in a red giant star \citep{Frandsen02}. 
Subsequently, researchers have seen evidence of 
oscillations in about a dozen K and late G type giants, both in nearby
field stars \citep{Ridder06,Barban07,Tarrant07} and
members of the open cluster \objectname[M67]{M67} \citep{Stello07}.

However, 
it remains uncertain if
accurate mode frequencies are possible to obtain, since observational
indications are that mode lifetimes could be too short \citep{Stello06}, in
disagreement with theory \citep{HoudekGough02}. Further, these stars might
pulsate predominantly in radial mode overtones as suggested by
\citet{Dalsgaard04} or indeed show additional non-radial pulsations as
indicated by \citet{Hekker06} and \citet{Kallinger07}. Without a full
understanding of 
which modes we observe, we cannot exploit the full potential of the
asteroseismic analysis. 

In this paper we aim at utilizing the characteristic acoustic frequency,
denoted \numax. This asteroseismic quantity
is relatively insensitive to the mode lifetime and the number of
excited modes, but still enables us to obtain information about the stellar
interiors. 
In particular, we will use \numax\  to estimate the masses of a
sample of K~giants observed  
with the star tracker on the {\it Wide-Field Infrared Explorer} (\wire)
satellite. 

\section{Observations and data reduction}\label{observations}
\begin{figure}
\plotone{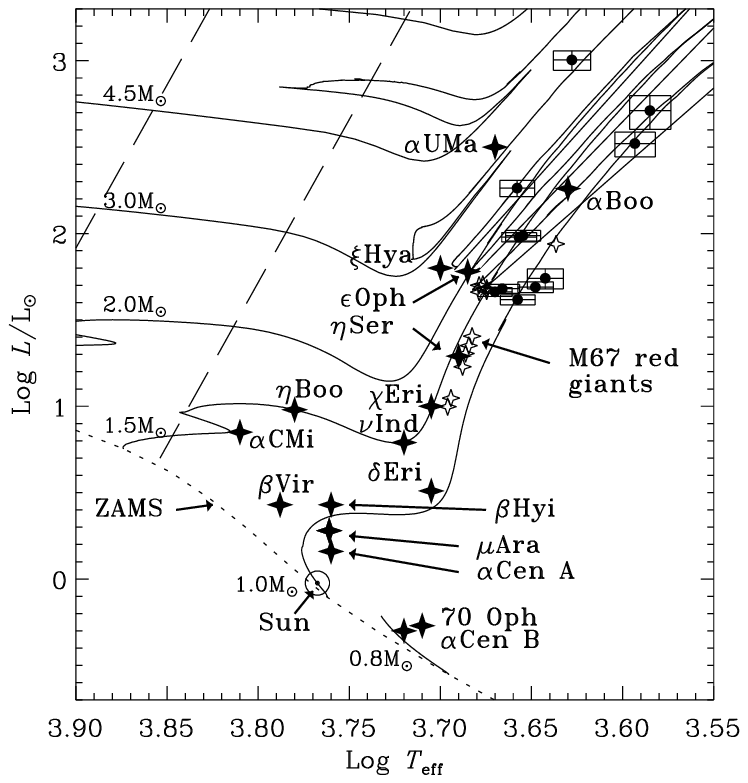}
\caption{H-R diagram with our 11 target stars including
  1$\sigma$-error boxes. Additional bright field stars (filled star
  symbols) and M67 
  cluster members (empty star symbols) that show evidence of solar-like
  oscillations are marked. Dashed lines indicate the approximate location 
  of the classical instability strip. Solid lines are evolution tracks. \label{fig1}} 
\end{figure}

During its lifetime \wire\ observed more than 40 evolved stars,
of which we have selected a subset of 
11 K~giants that
all have long time series and good noise characteristics.
The time series span from 15 to 61 days, and the observations were obtained
from February 2004 to May 2006. 
In Fig.~\ref{fig1} we show the location of the giants in the H-R diagram,
together with stars that have previously been reported to show evidence of
solar-like oscillations. We extracted the complete set of evolution tracks
($M=0.5$--$10\,$M$_{\odot}$) 
from the BaSTI database \citep{Pietrinferni04}, based on their
alpha-enhanced standard solar models without overshooting (Z=0.0198,
Y=0.273). 
We only plot a representative subset of tracks in
Fig.~\ref{fig1}. 


We obtain the raw light curves 
using the pipeline described by \citet{Bruntt05}.
Data reduction proceeds in two stages. First, we remove obviously 
aberrant points from the time series. This is done by examination of
instrumental magnitude, FWHM of the stellar profile, centroid position, and 
background level as a function of time. Generally, the majority of data
points examined in this way fall into a well-defined group with a small
percentage of outliers, which are removed. 
We next remove effects from high levels of scattered light at the beginning
of each orbit by phasing each time series at the satellite orbital period, 
and then subtract a smoothed version of the phased light curve.
The resulting mean-subtracted time series is used for the analysis
described below.  

The \wire\  star tracker obtained images of each star with a  
cadence of $2\,$Hz, providing a few million observations per
star. However, with the expected long oscillation period of the 
K~giants ($P>6$ hours) we are able to speed up the Fourier analysis
(see Sect.~\ref{analysis}) by binning the data in intervals equal to the
satellite orbital period ($\sim$90 minutes), corresponding  
to 230--830 data points per star. The resulting time series have a
continuous sampling of one data point per orbit with only very few
small gaps, which translates into a Nyquist frequency of $90\,$\muhz.

\section{Stellar parameters}
To facilitate our investigation we have derived the stellar parameters for
each star. This enables us to predict the characteristic acoustic
frequency, \numax$_{,\mathrm{pre}}$, 
where we would
expect to see excess power in the Fourier spectrum. The stellar parameters
are listed in Table~\ref{tab1}, and sorted according to
\numax$_{,\mathrm{obs}}$, which is the frequency of the observed excess
power (see Sect.~\ref{analysis}). 

In the following we explain each column of Table~\ref{tab1}. The $V$
magnitude is obtained from   
the \mbox{SIMBAD} database, derived as the mean of the listed values from up to
six sources, and the standard deviation is adopted as a conservative
uncertainty. The infrared $K$ band magnitude is taken from 2MASS
\citep{Cutri03}, the parallax, $\pi$, is from the new Hipparcos release 
\citep{Leeuwen07}, and the effective temperature is derived using the
$T_{\mathrm{eff}}$-$(V-K)$ relation by \citet{Alonso99}. The internal error
of the $T_{\mathrm{eff}}$-$(V-K)$ relation is only $25\,$K. However, we adopt 
$\sigma_{T_{\mathrm{eff}}}=100\,$K as a realistic uncertainty, in agreement
with \citet{Kucinskas05}. 
The color-temperature relation requires as input [Fe/H], but is
insensitive to log$\,g$. 
For five stars in our sample, spectroscopic information is
available within a uniform collection \citep{McWilliam90}, and they all
agree with having solar metallicity. Hence, we assume solar metallicity for
all our targets, which is reasonable for such nearby stars. The
spectroscopically determined effective 
temperatures are in good agreement with those derived in Table~\ref{tab1}. 
Three stars have interferometrically calibrated $T_{\mathrm{eff}}$
\citep{Blackwell98,Benedetto98}, which also agree with our quoted values.

We then derive the luminosity from 
$\mathrm{log}(L/\mathrm{L}_{\odot})=-[V+BC-5\mathrm{log}(1000/\pi)+5-\mathrm{M}_{\mathrm{bol,\odot}}]/2.5$,
where $BC$ is the bolometric correction obtained from the $BC$-$(V-K)$
relation by \citet{Alonso99}, and $\mathrm{M}_{\mathrm{bol,\odot}}=4.75$
is the solar absolute bolometric magnitude (recommendation of IAU 1999).
Since these are all nearby stars we neglect interstellar absorption, $A_V$,
but we do include $\sigma_{A_V}$ in the error budget. 
In general, the relative contributions to $\sigma_{L}$ from the individual
uncertainties are
$\sigma_{BC}>\sigma_{\mathrm{A_V}}>\sigma_{\pi}>\sigma_{V}$,
with $\sigma_{BC}$ being dominated by $\sigma_{T_{\mathrm{eff}}}$, while
$\sigma_{\pi}$ and $\sigma_{V}$ are 
negligible for most stars. 

The quoted range in the photometric mass, $M_{\mathrm{phot}}$(see Table~\ref{tab1}), corresponds to the lowest and
highest masses of all tracks that go through the $1\sigma$-error box in
the H-R diagram, while taking mass loss into account in the stellar models
(mass loss parameter $\eta=0.4$; see \citet{Pietrinferni04} and references
herein).
We note that this approach will always underestimate the true mass range
within the $1\sigma$-error box due to the non-zero mass step, $\Delta M$, in  
our grid of tracks, which is $\Delta M=0.1\,$M$_{\odot}$ for
$0.9<M/$M$_{\odot}<2.0$, $\Delta M=0.2\,$M$_{\odot}$ for
$2.0<M/$M$_{\odot}<3.0$, and $\Delta M=0.5\,$M$_{\odot}$ for 
$3.0<M/$M$_{\odot}<10.0$. Hence, if a track is just outside the error box,
our quoted mass range will be smaller by almost one mass step. 
Within the age of the universe, none of the tracks with progenitor mass
$M\lesssim0.85\,$M$_{\odot}$ and solar metallicity have 
yet evolved to the red giant phase (see Fig.~\ref{fig1}). Hence, we use
the $0.90\,$M$_{\odot}$ track to define the minimum masses\footnote{We note
  that for sub-solar metallicities we would expect to see red giants with
  masses below $0.85\,$M$_{\odot}$.}. 
It is beyond the scope
of this Letter to go into detail about additional systematic errors in these
mass estimates, originating from metallicity, overshooting, and differences
in evolution codes.

\begin{figure}
\plotone{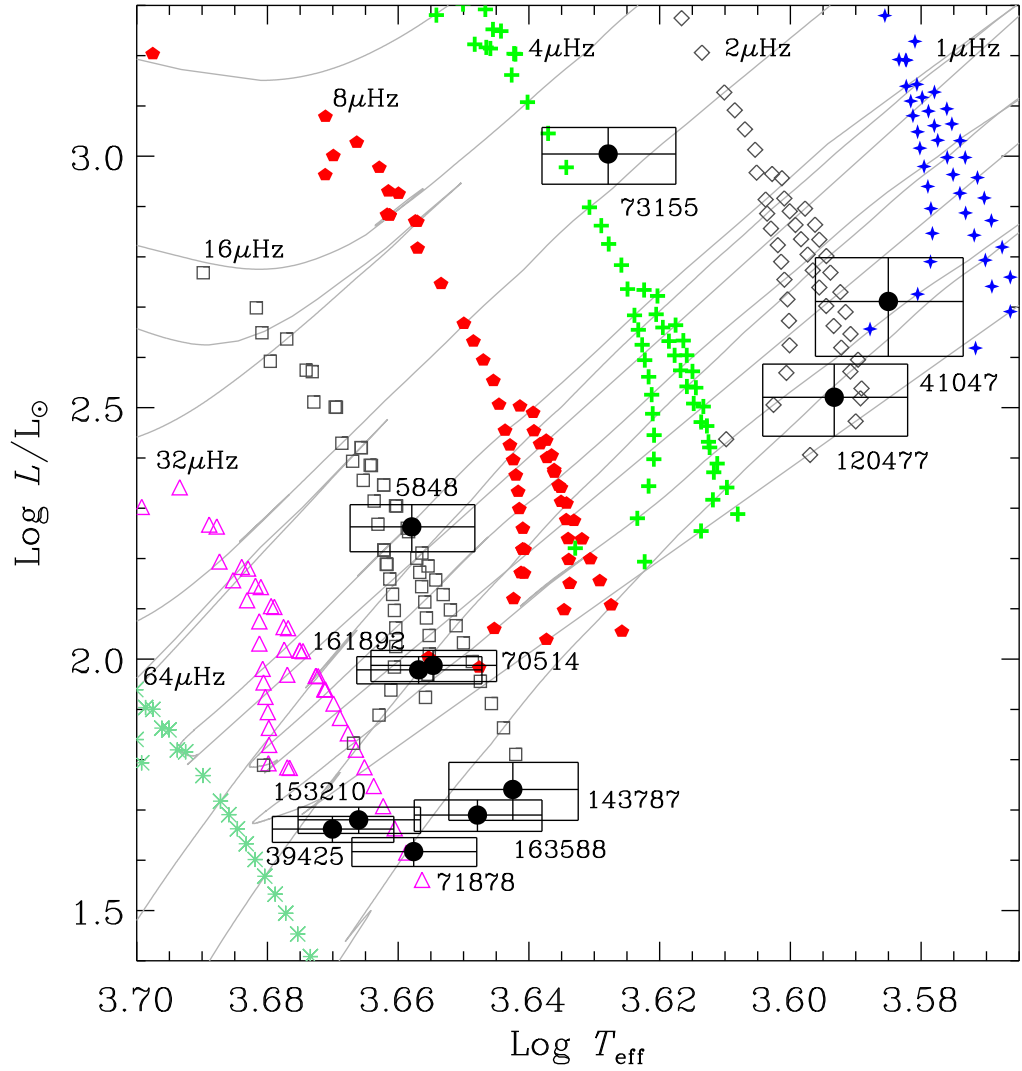}
\caption{H-R diagram of region around our target stars (solid dots inside
 $1\sigma$-error boxes).
 Gray lines are a representative sample of evolution tracks. 
 The additional symbols show points along the evolution tracks
 corresponding to a given value of \numax$_{,\mathrm{pre}}$ (values are
 indicated for each symbol). 
 \label{fig3}}  
\end{figure}

Finally, we calculate the frequency,
\numax$_{,\mathrm{pre}}$ (see Table~\ref{tab1}), where the highest excess
power from solar-like oscillations is expected in the Fourier
spectrum. This is obtained by scaling the acoustic cut-off frequency of the
Sun 
\citep{Brown91}  
\begin{equation}
 \nu_{\mathrm{max}}=\frac{M/\mathrm{M}_{\sun}\,(T_{\mathrm{eff}}/5777\,\mathrm{K})^{3.5}} 
                                 {L/\mathrm{L}_{\sun}}\, \nu_{\mathrm{max},\odot}\, ,
\label{eq_freq_max}
\end{equation}
\noindent 
where we use the solar value, $\nu_{\mathrm{max},\odot}=3021\pm 27$ \muhz,
found from 10 independent 30-day time series from the 
VIRGO instrument on board the SOHO spacecraft, 
using the same approach as for the
K~giants (see Sect.~\ref{analysis}).
Equation~\ref{eq_freq_max} has been shown to give very good estimates of
the frequency of maximum oscillation power based on observations of mostly
less evolved stars with relatively well constrained 
masses \citep{BeddingKjeldsen03} compared to our targets. 
As with $M_{\mathrm{phot}}$, we also
quote a range for \numax$_{\mathrm{,pre}}$, which 
states the
extreme values of \numax$_{\mathrm{,pre}}$ that are within the
$1\sigma$-error box in the H-R diagram. 
To illustrate this we show an H-R diagram close-up of our target stars in
Fig.~\ref{fig3}. The 
black dots correspond to the stellar parameters given in Table~\ref{tab1}
(heading: ``Derived''), and the $1\sigma$-error boxes and HD numbers are
also shown.    
We plot $L$, $T_{\mathrm{eff}}$ for selected values of
\numax$_{\mathrm{,pre}}$ along each evolution track, which clearly shows the
complexity of estimating \numax\  
from the stellar position in this part of the H-R diagram. 
Note that, similar to Fig.~\ref{fig1}, only a selected sample of all
evolution tracks are plotted. 
Along each evolutionary track, a given value of \numax$_{\mathrm{,pre}}$
(shown as identical symbols) can occur up to three times, twice on the red 
giant branch (ascending and descending), and once while ascending the
asymptotic giant branch.
In any region where the tracks cross, the mass, and hence \numax, is not
uniquely determined by the location in the H-R diagram.
The last two columns in Table~\ref{tab1} 
will be explained in the following section.

\section{Asteroseismic analysis}\label{analysis}
\begin{figure}
\epsscale{.35}
\plotone{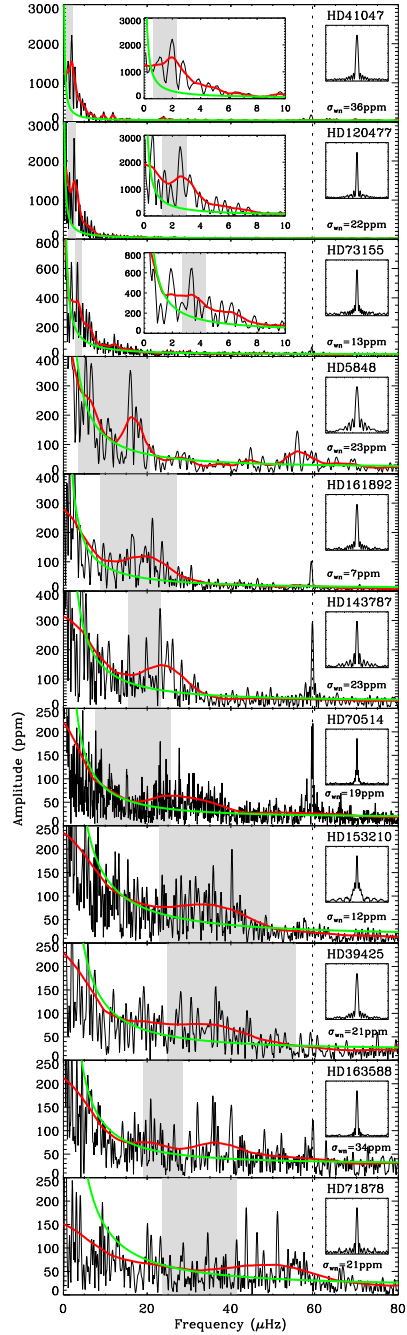}
\caption{Amplitude spectra of the 11 K~giants (HD numbers are
  indicated). Note the increasing amplitude scale on the ordinate. In each
  panel the red solid   
  line is a smoothed version of the spectrum and the green monotonically
  decreasing line is a fit to the noise. Gray shaded areas indicate the 
  frequency interval of \numax$_{,\mathrm{pre}}$. The dotted line marks one
  third of the \wire\ orbital frequency, and $\sigma_{\mathrm{wn}}$ is the
  square root of the mean power in the range: 70--80$\,$\muhz.
  The inset is the spectral window in amplitude, shown in the same
  horizontal scale as the main panels.  \label{fig2}}  
\end{figure}

To look for evidence of solar-like oscillations, we first calculate the
Fourier spectrum of each star, which is shown in amplitude
(=$\sqrt{\mathrm{power}}$) in Fig.~\ref{fig2}.  
The monotonic green curve is a fit to the noise, 
described by 
\mbox{$\sigma(\nu)=a/\nu^{2}+\sigma^2_{\mathrm{wn}}$}
in power. 
The parameters $a$ and $\sigma^2_{\mathrm{wn}}$ are determined 
following the approach by \citet{Stello07}. 
In addition, we smooth each spectrum (red
curve) to remove the detailed structure of the excess power. Smoothing was
done using a moving box average twice with a width equal to twice the expected
frequency separation of adjacent radial modes derived as:
$\Delta\nu=(M/$M$_{\odot})^{0.5}(R/$R$_{\odot})^{-1.5}134.92\,$
\muhz~\citep{KjeldsenBedding95}. We note that the location of the excess
hump in the smoothed spectrum does not depend
strongly on the mass adopted in the calculation of $\Delta\nu$. 

From Fig.~\ref{fig2} we see a clear trend of excess power shifting to
higher frequency from top to bottom. A power excess at higher frequencies
generally corresponds to less luminous stars, but this trend is 
modulated by mass and temperature, 
in agreement with Eq.~\ref{eq_freq_max}.
The gray shaded areas in each panel
indicate the predicted region of 
the excess power (Table~\ref{tab1}, Col.8).
We further note that the amplitude seems to decrease as the width of
the envelope increases. Similar trends can be seen in 
\citet{Kjeldsen05} for less luminous stars. 

Despite the large uncertainty in the stellar masses,
$M_{\mathrm{phot}}$, and hence in \numax$_{,\mathrm{pre}}$, 
our results in Fig.~\ref{fig2} confirm that Eq.~\ref{eq_freq_max} is
valid for K~giants in a large luminosity range.
Now, if we turn the argument around, assuming that Eq.~\ref{eq_freq_max} is
exact, and hence interpret any observed deviation from this relation to be
largely due to an inaccurate ``photometric'' mass, we can use it to
infer an ``asteroseismic'' mass. 
To obtain the asteroseismic mass we measure \numax\  by first subtracting
the noise fit from the smoothed spectra, and then we locate the maximum
of the residual (see Table~\ref{tab1}, Col.9). 
By smoothing we obtain a more robust measure than trying to locate the
strongest oscillation mode. 
To estimate the uncertainty 
of \numax$_{\mathrm{,obs}}$, we make 20 simulations of each time series
following the approach by \citet{Stello04}. The simulations show that
\numax$_{\mathrm{,obs}}$ has an uncertainty of roughly 10\%, but it varies
somewhat from star to star, due to 
differences in pulsation characteristics, the duration of the time series,
and the noise level.
We found that our results do not depend
strongly on the adopted mode lifetime, $\tau$, in the range $3\,$d $<\tau<
20\,$d, that we investigated. 
We then finally determine $M_{\mathrm{seis}}$ using
Eq.~\ref{eq_freq_max}. 

Our results show that for stars located in the region
where evolution tracks cross, $M_{\mathrm{seis}}$ has a lower
$1\sigma$ uncertainty than the $M_{\mathrm{phot}}$
mass range (assuming Eq.~\ref{eq_freq_max} is exact). 
In regions without crossings 
the benefit from having measured \numax$_{\mathrm{,obs}}$ is less
obvious. For those stars, we need longer time series to obtain a lower
uncertainty in \numax$_{\mathrm{,obs}}$, which generally dominates the
uncertainty on our mass estimate. 
For a few stars we are able to measure the large frequency separation, $\Delta\nu$, 
which potentially can give a more precise mass estimate than \numax. 
However, for
our present data \numax\ provides the smallest mass uncertainties.



\section{Conclusions}
We have analyzed photometric time series of 11 nearby K~giants 
obtained 
with the \wire\ satellite. The Fourier transforms show clear evidence that
oscillation power shifts to higher frequencies for less luminous stars, as
anticipated from scaling the solar frequency of maximum power,
\numax$_{,\odot}$.
We were able to measure \numax\  and made simulations of each star to obtain
a realistic uncertainty of the measurement. Using a simple scaling
relation, which relates this frequency to the stellar parameters
$T_{\mathrm{eff}}$, $L/\mathrm{L}_{\odot}$, and $M/$M$_{\odot}$, we 
estimated an asteroseismic mass. For several stars this
approach provides a
significantly lower uncertainty of the mass relative to the
classical mass estimate based purely on comparing stellar evolution tracks
with the location in the H-R diagram. 

These results show exciting prospects for the current COROT mission
\citep{Baglin98} and the upcoming  
Kepler satellite \citep{Dalsgaard07}, which will both provide much more
extended times series than \wire. With longer time series we can  
potentially acquire
lower uncertainties in the \numax\ 
measurements, and hence  
more precise mass estimates. This will indeed be possible with Kepler,
which will obtain parallaxes, and hence luminosities, of its target stars.
Our approach 
could be particularly valuable in cases where 
the oscillation power does not allow accurate detection of 
the large frequency separation, $\Delta\nu$, 
due to an insufficient number of modes with high
signal-to-noise.
This might, in fact, include most faint solar-like pulsators, as well as
bright giant stars if indeed the mode lifetime does not increase with
increasing oscillation periods \citep{Stello06}.

\acknowledgments
We acknowledge the financial from 
the ARC, DASC, and FNU.
This research
has made use of the SIMBAD database, operated at CDS, Strasbourg, France.
We thank Tim Bedding to comments on the manuscript.

\clearpage

\bibliography{bib_complete}

\clearpage

\input{tab1}

\clearpage

\clearpage

\clearpage

\clearpage

\end{document}

%% file: tab1.tex
\begin{table}
{\footnotesize
\begin{center}
\caption{Stellar parameters of the \wire\ K~giants.\label{tab1}}
\begin{tabular}{r|r@{(}lrr|rr@{(}l|rr|rr}
\tableline\tableline
       & \multicolumn{4}{c}{Literature}   & \multicolumn{3}{c}{Derived}                               & \multicolumn{2}{c}{Evolution tracks}                     & \multicolumn{2}{c}{Asteroseismology}                      \\
HD                            & \multicolumn{2}{c}{$V$}   & \multicolumn{1}{c}{$V-K$}     & \multicolumn{1}{c}{$\pi$}     & \multicolumn{1}{c}{$T_{\mathrm{eff}}$}   & \multicolumn{2}{c|}{$L/$L$_{\odot}$} & \multicolumn{1}{c}{$M_{\mathrm{phot}}/$M$_{\odot}$} & \multicolumn{1}{c}{\numax$_{\mathrm{,pre}}$}& \multicolumn{1}{c}{\numax$_{\mathrm{,obs}}$} & \multicolumn{1}{c}{$M_{\mathrm{seis}}/$M$_{\odot}$}\\
       & \multicolumn{2}{c}{(mag)}     & \multicolumn{1}{c}{(mag)}     & \multicolumn{1}{c}{(mas)}     & \multicolumn{1}{c}{(K)\tablenotemark{a}} & \multicolumn{2}{c|}{}                &                                 & \multicolumn{1}{c}{(\muhz)}                 & \multicolumn{1}{c}{(\muhz)}                   &                                \\
\tableline

\objectname[HD41047]{41047}   & 5.543&05)\tablenotemark{b} & 3.80(22)  &  5.30(33) & 3846                 &  514&114)                            &  0.60--1.60                     & 0.7--2.3                & 1.99(27)                   & 1.40(34)                       \\
\objectname[HD120477]{120477} & 4.049&19) & 3.61(18)  & 12.37(23) & 3920                 &  332&54)                            &  0.61--1.18                     & 1.3--3.0                & 2.61(63)                   & 1.11(33)                       \\
\objectname[HD73155]{73155}   & 5.008&05) & 2.99(27)  &  3.83(19) & 4245                 & 1010&131)                           &  2.78--3.98                     & 2.7--4.4                & 3.39(99)                   & 3.3(1.1)                       \\
\objectname[HD5848]{5848}     & 4.225&29) & 2.57(24)  & 11.63(15) & 4549                 &  183&20)                            &  1.37--2.60                     & 3.6--20.7               & 16.3(1.5)                  & 2.27(41)                       \\
\objectname[HD161892]{161892} & 3.203&06) & 2.58(27)  & 25.91(15) & 4538                 &   95&6)                             &  0.62--1.79                     & 8.8--27.0               & 19.7(2.2)                  & 1.44(21)                       \\
\objectname[HD143787]{143787} & 4.985&35) & 2.77(27)  & 15.71(31) & 4390                 &   55&7)                             &  0.89--0.99                     & 15.4--23.2              & 23.3(1.6)                  & 1.11(19)                       \\
\objectname[HD70514]{70514}   & 5.060&08) & 2.61(25)  & 10.98(16) & 4515                 &   97&7)                             &  0.62--1.79                     & 7.7--25.7               & 24.7(2.8)                  & 1.88(29)                       \\
\objectname[HD153210]{153210} & 3.197&08) & 2.47(18)  & 35.65(20) & 4635                 &   48&3)                             &  0.78--1.49                     & 22.9--49.3              & 34.8(2.5)                  & 1.19(14)                       \\
\objectname[HD39425]{39425}   & 3.113&09) & 2.42(34)  & 37.42(12) & 4678                 &   46&3)                             &  0.77--1.59                     & 24.8--55.6              & 34.8(5.3)                  & 1.10(20)                       \\
\objectname[HD163588]{163588} & 3.743&10) & 2.70(19)  & 28.98(12) & 4445                 &   49&4)                             &  0.89--1.09                     & 19.1--28.4              & 35.9(2.0)                  & 1.45(17)                       \\
\objectname[HD71878]{71878}   & 3.760&10) & 2.57(30)  & 30.32(10) & 4546                 &   41&3)                             &  0.89--1.19                     & 23.5--41.3              & 51.1(3.9)                  & 1.62(20)                       \\
\tableline
\end{tabular}
\tablenotetext{a}{The adopted $1\sigma$ uncertainty of $T_{\mathrm{eff}}$
  is $100\,$K.}
\tablenotetext{b}{Numbers in parentheses are uncertainties, e.g. for 
  HD41047 the $V$ magnitude and its uncertainty is $5.543\pm0.005\,$mag.}
\end{center}
}
\end{table}